\documentclass[useAMS,usenatbib]{mn2e_astroph}

\usepackage{graphicx}

\def\nct#1{\nocite{#1}}

\title[Artefacts of cartography in pulsar polarization]
{Artefacts of circumpolar cartography in radio pulsar polarization}

\author[J.~Dyks]
{J.~Dyks
\\
Nicolaus Copernicus Astronomical Center, Rabia\'nska 8, 87-100, Toru\'n,
Poland\\
}
\begin{document}

\date{Accepted 2020 Apr 20. Received 2020 Feb 6; in original form 2020 Jan 31}


\maketitle

\label{firstpage}

\begin{abstract}
Single pulse data on radio pulsar polarization are traditionally presented
in the form of two-dimentional greyscale patterns with the pulse longitude and polarization
angle (PA) on the horizontal and vertical axis, respectively. 
Such diagrams reveal several enigmatic polarization effects: 1) bifurcations
and loops of PA curve under central pulse components, 2) vertical spread of flux
at all PA values, 
3) exchange of power content 
between PA tracks of two orthogonal
polarization modes (OPMs), and 4) peripherically-flat PA swings that span
more than 180 degrees. It is shown that all these phenomena 
result 
from passage of observed polarization state near the pure-V pole of Poincar{\'e} sphere. 
Much of their complexity results from cartographic 
transformation from Poincar{\'e} sphere to the longitude-PA plane. 
An odd number of near-pole passage produces apparent replacement of OPMs
power in the profile wings, although the same amount of flux keeps staying in each modal patch on
the Poincar{\'e} sphere. The fitting of pulsar PA curves should therefore allow
for transitions between the primary (strong) and secondary (weak) PA track. 
The Stokes-space (or Poincar{\'e}-sphere) representation of pulsar polarization
data contains crucial polarization information and needs to accompany the
traditional viewing if the published figures are to be fully useful for 
interpretation.
\end{abstract}


\def\lap{\hbox{\hspace{4.3mm}}
         \raise1.5pt \vbox{\moveleft9pt\hbox{$<$}}
         \lower1.5pt \vbox{\moveleft9pt\hbox{$\sim$ }}
         \hbox{\hskip 0.02mm}}

\def\rwobs{R_W}
\def\rwcon{R_W}
\def\rwstr{R_W}
\def\winobs{W_{\rm in}}
\def\woutobs{W_{\rm out}}
\def\phm{\phi_m}
\def\phmi{\phi_{m, i}}
\def\thm{\theta_m}
\def\dres{\Delta\phi_{\rm res}}
\def\win{W_{\rm in}}
\def\wout{W_{\rm out}}
\def\rin{\rho_{\rm in}}
\def\rout{\rho_{\rm out}}
\def\phin{\phi_{\rm in}}
\def\phout{\phi_{\rm out}}
\def\xin{x_{\rm in}}
\def\xout{x_{\rm out}}

\def\thmin{\theta_{\rm min}^{\thinspace m}}
\def\thmax{\theta_{\rm max}^{\thinspace m}}

\section{Introduction}
\label{intro}

Radio pulsar polarization has been studied both 
observationally (van Straten \& Tiburzi 2017; Dai et al.~2015; Noutsos et
al.~2015; McKinnon \& Stinebring 1998; Karastergiou 2009) 
\nct{vst17, dhm15, nsk15, ms98, k2009} and theoretically (Hakobyan et al.~2017, Wang et
al.~2010; Petrova \& Lyubarskii 2000) 
\nct{hbp17, wlh10, pl00}
for about half a century.
Direct interpretation of polarization angle (PA) in terms of the sky-projected
magnetic field direction justifies the usual way of data representation,
which is the two-dimentional histogram (greyscale plot) of single-pulse samples on the
plane of pulse longitude and PA (hereafter longitude-PA plane) (Rankin \&
Rathnasree 1997; 
Mitra et al.~2015; 
Stinebring et al.~1984; Hankins \& Rankin 2010). 
\nct{hr10, mar2015, rr97, scr84}
Another presentation of pulsar polarization, namely the projection on the
Poincar{\'e} sphere, has also been used though less frequently (Edwards \&
Stappers 2004; Os{\l}owski et
al. 2014). \nct{ovb14, es04}

The traditional presentation method (on the longitude-PA plane) resulted in
identification of several enigmatic polarization effects. 1) The PA curve
has been found to undergo bifurcations and complex distortions, usually observed under the
central pulse component (`core') (eg.~B1237$+$25, Smith et al.~2013). \nct{srm13} 
2) At some pulse longitudes, especially
near transitions between orthogonal polarization modes (OPMs) the observed flux is 
spread across all PA values. According to Ilie (2019, p.~129)\footnote{The
PhD thesis of Cristina Ilie is freely available at the University of Manchester:
https:$//$www.research.manchester.ac.uk$/$ portal$/$files$/$104260195$/$FULL$_{-}$\negthinspace TEXT.PDF 
}   
\nct{i19}
 the radiation at OPM transitions in PSR B1451$-$68 (J1456$-$6843) 
 is almost entirely circularly polarized, despite the average polarization degree is
$\sim\negthinspace10\%$ because of the integration over two OPMs. Average profiles of other
pulsars show even much higher levels of circular polarization degree: 
$V/I$ exceeds $30\%$ in B1913$+$16 (Weisberg \& Taylor 2002; \nct{wt02}
see Fig.~1 in Dyks 2017), \nct{d2017} $40\%$ in the loop-like PA distortion observed in 
B1933$+$16 (Mitra et al.~2016). \nct{mra2016}
In the abnormal pulsation mode of PSR
B1237$+$25 the average $V/I$ reaches almost $30\%$ (Srostlik \& Rankin 2005).
\nct{sr05}
These values are large 
despite the averaging over the usual intrinsic spread of polarization states
in different rotation periods, 
and despite the likely cancelling from summation over two orthogonal polarization
modes. Such high polarization degree cannot result from  
   noncoherent superposition of near-equal OPMs which is the most likely cause  
in the case of a weakly polarized signal (Melrose et al.~2006). 
\nct{mmk2006}
3) in some objects (eg.~PSR B1857$-$26, Mitra \& Rankin 2008) \nct{mr2008} the power of OPMs 
seems to become exchanged,
in the sense that the primary (stronger) polarization mode starts to follow
the PA track of the secondary (weaker) mode. The distribution of polarized power observed in the outer wings
of such profiles looks therefore asymmetric, with most power in the leading wing following
one PA track, while most of the trailing wing radiation follows the other (orthogonal)
track. 4) In contrast to prediction of the rotating vector model (RVM) 
the range of PA variation within some profiles (eg.~PSR B1857$-$26, Mitra \&
Rankin 2008) \nct{mr2008} exceeds
$180^\circ$, despite the horizontal orientation of the PA curve 
in the profile wings suggests equatorward viewing geometry. 

\section{Quasi-meridional circularization 
or near-vertical motion through Poincar{\'e} sphere}

\begin{figure}
\includegraphics[width=0.48\textwidth]{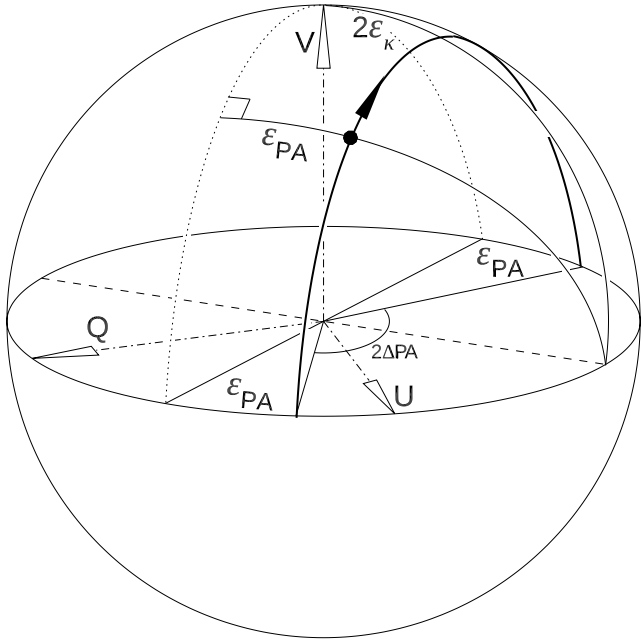}
\caption{Geometry of the near meridional circularization on Poincar\'e
sphere. The polarization state (bullet) rotates around the linear retarder
axis (dashed), following the thick solid arc.  The dotted meridian contains the V pole 
on top of the sphere. $\epsilon_{PA}$ represents the discrepancy of observed PA
difference from orthogonality, whereas $\epsilon_\kappa$ is the difference of
extreme ellipticity
angle $|\kappa|$ from $45^\circ$.
}
\label{vpp}
\end{figure}

 High circular polarization fraction $|V|/I$ has been 
observed in several pulsars, eg.~in the PA loop of B1933$+$16 
(Mitra et al.~2016). Empirical models of this PA distortion 
(eg.~Fig.~13 in Dyks 2017, Fig.~23 in Dyks 2019) assume coherent
summation of orthogonally-polarized waves with near equal amplitude and
with longitude-dependent phase lag between the waves. The increasing phase lag 
corresponds to the near-meridional rotation of polarization state 
on the Poincar{\'e} sphere (the usual action of a linear retarder).
This involves the passage of the
polarization state near the pure-V pole (hereafter V pole) 
 of the Poincar{\'e} sphere. 
 
Moreover, new single-pulse data (Ilie 2019) clearly show that polarization
state of pulsars sometimes changes with pulse longitude 
in a rather special way: the mode-specific patch of flux on
the Poincar{\'e} sphere moves vertically, or quasi-vertically upward or
downwards. In the course of its motion, the modal patch is passing very close to
the V pole of the Poincar{\'e} sphere.
An example of such phenomenon is PSR B1451$-$68, in which both orthogonal 
modal patches undergo such circularization and pass across both V 
poles (Fig.~3.45 in Ilie 2019). \nct{i19}
There are in total four near-pole passages in this object, 
therefore, the eccentricty angle of polarization ellipse follows two
curves that form a
bow-tie shape (Fig.~3.44 in Ilie 2019). \nct{i19} 

The physical reason for the circularization is unknown, however, it is not needed to
understand the enigmatic look of the above-mentioned effects, at least on geometrical level. 
The meridional rotation of the polarization state may result from intrinsic
properties of the emission mechanism or from propagation effects. 
To facilitate the discussion, however, 
it is assumed in the following that the circularization
results from coherent superposition of two orthogonally
polarized natural waves (Dyks 2017;
2019; Edwards \& Stappers 2004). \nct{d2017, d2019, es04} 
If the phase delay between the combined 
waves is changing with
pulse longitude, the observed flux patch moves on the Poincar{\'e} sphere: it is rotated around
an axis defined by the
natural modes. 
The patch observed on
the sphere can stay local (when it is identified as the `observed polarization
mode') or can extend into an arc in the case of spread in the
phase delay values (which also has been observed, eg.~near the V poles in
B1451$-$68). 
If the natural waves are linearly polarized the axis of patch rotation is
equatorial, ie.~it is
contained in the equatorial plane (dashed lin
e in Fig.~\ref{vpp}). 
The ratio of the waves' amplitudes $E_2/E_1$ 
(or the mixing angle $\psi=\arctan{[E_2/E_1]}$) then determines whether the
observed patch follows a meridian or a small circle on the sphere. 


Let us assume that only one polarization mode is observed, ie.~the supposed
natural waves' superposition produces only one modal patch on the Poincar{\'e}
sphere. In the case of identical
amplitudes of the superposed waves, the patch moves meridionally, passing
centrally through the pole (this would correspond to the motion along the
dotted meridian in Fig.~\ref{vpp}).
If the amplitudes are similar but not identical,  
the patch moves nearly
meridionally, and the observed PA stays almost constant until the patch is passing
near the V pole of the sphere. In Fig.~\ref{vpp} such type of motion is shown
with the bullet which is following the solid non-meridional circle.
 During the near-pole passage, the closest
approach angle of $2\epsilon_{\kappa}$ is reached, and the PA quickly
changes to an almost orthogonal value. The discrepancy from the
orthogonality of the initial and final PA can be denoted as 
$\epsilon_{\rm PA}=90^\circ-\Delta PA$. 
If the traverse of the patch starts and ends
near the
equator, then $\epsilon_{\rm PA}$ is equal to the minimum angular distance
from the vertical (Stokes V) axis :
\begin{equation}
\epsilon_{\rm PA} = 90^\circ - 2|\kappa_{\rm max}|=2(45^\circ-|\kappa_{\rm
max}|) \equiv 2\epsilon_\kappa,
\label{epa}
\end{equation}
where $\kappa_{\rm max}$ is the
maximal (or minimal) ellipticity angle, i.e.~$\kappa=0.5\arctan(V/L)$, where
$L$ is the linearly polarized flux. 
Eq.~(\ref{epa}) is valid if other effects (eg.~of the RVM origin) do not
contribute to the change of PA. In flat parts of an observed PA curve,
which often occur in millisecond pulsars, and sometimes in normal pulsars, this equation can be used to test
if the superposed natural waves are linearly polarized (i.e.~if the patch rotates
around an equatorial axis). In general it is necessary  
 to take into account the RVM 
changes of PA, and the off-equatorial orientation of the patch-rotation axis.

\section{Cartographic effects}
\subsection{Apparent OPM jumps from passage through the V pole}

\begin{figure}
\includegraphics[width=0.48\textwidth]{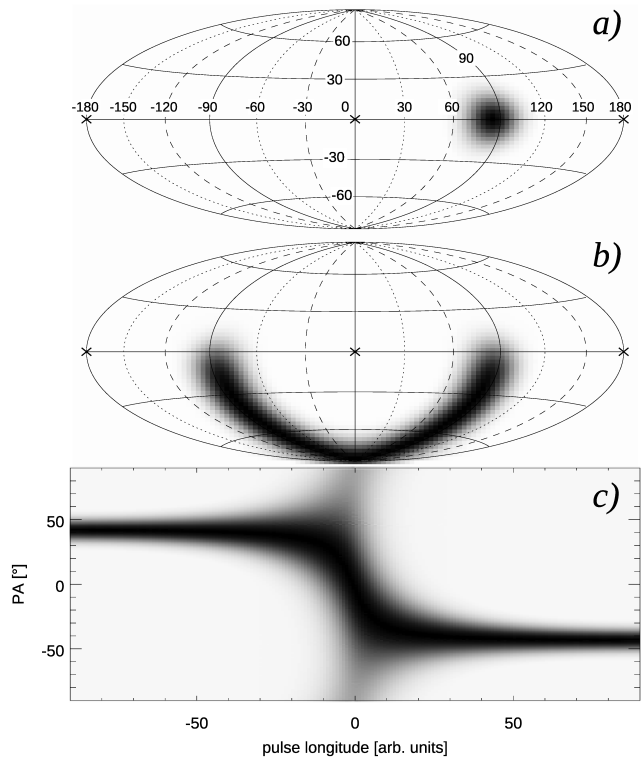}
\caption{Transformation of flux patch motion from the Poincar{\'e} sphere to the
longitude-PA diagram. {\bf a)} Initial patch position viewed in the Hammer equal
area projection. {\bf b)} Trace of the patch motion through the sphere. The
patch was uniformly rotated around the axis piercing the `x' symbols.
{\bf c)} PA track that corresponds to the motion shown in panel b. 
The patch was a Gaussian with $\sigma=10^\circ$. The closest approach to the V pole
was $5^\circ$. Pulse longitude is set equal to the phase delay
(patch rotation angle). Numbers in a) give azimuth and latitude in
degrees. The two `x' points at azimuth $\pm180^\circ$ correspond to the same
point on the sphere.
}
\label{blobth}
\end{figure}
\begin{figure}
\includegraphics[width=0.48\textwidth]{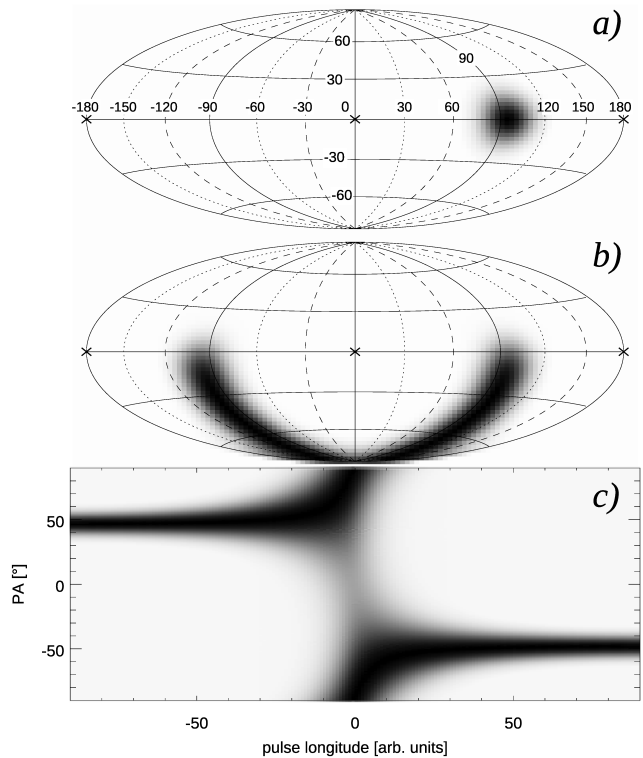}
\caption{Same as in Fig.~\ref{blobth} except now the patch is passing on the
other side of the pole. The opposite turn of PA track (in comparison to
Fig.~\ref{blobth}c)
is observed as PA bifurcations under core components (eg.~in PSR B1237$+$25).
}
\label{blobth2}
\end{figure}

Opposite azimuths mean orthogonal PAs, therefore, 
the passage of the patch (ie.~passage of the observed polarization state) through the
V pole of the Poincar{\'e} sphere 
results in an \emph{apparent} OPM transition: the radiative power moves to the
other (orthogonal)
PA track, although there is no real OPM change, because only one patch of
flux may actually exist on the whole Poincar{\'e} sphere. 

If the secondary (weaker) orthogonal mode is also observed, another weaker patch
is present at the antipodal point of the Poincar{\'e} sphere. This secondary patch follows a
`mirrored'
motion on the sphere, ie.~the patch is passing through the other V pole 
and makes simultaneous transition to the other (originally
primary) PA track. 

The observed replacement of power between the primary and secondary PA
tracks is therefore only an apparent effect: the OPM transition between the
PA tracks is observed, however, it has nothing to do with one mode becoming
stronger than the other. The modes keep maintaining their typical power, it is their
location with respect to the V pole, not their power, that has changed. 

Such apparent OPM transition (effect of the near-pole passage) can possibly 
occur either odd
or even number of times in
a pulse profile, hence it is possible to observe the asymmetrically
polarized profiles with the primary (large) radiative flux 
on the leading side, and the secondary
(low) flux on the trailing side of the same (single) RVM track.  It must
be emphasized, however, that the
asymmetry can also be caused by a single usual (standard) OPM change.

In the case of a slightly non-meridional patch motion, the pole passage may result in
 a quasi-orthogonal
OPM jump. 
If the
patch is rotated around an equatorial axis (linearly polarized natural
waves, Fig.~\ref{vpp}), then the
PA will change by the amount given by eq.~1. This may be responsible for at
least some quasi-orthogonal OPM jumps observed in several objects. In other
cases the `real' modes (local patches on Poincar{\'e} sphere) are indeed
quasi-orthogonal (eg.~Ilie et al.~2020). \nct{iwj19}

\subsection{Complex polarization in the core components}

Because of the cartographic  transformation between the Poincar{\'e} sphere
and the longitude-PA plane, the innocuous near-pole passage of polarization
state causes extremely strong effects to appear in the longitude-PA diagrams. 
In particular, the complex polarization observed in pulsar cores (especially
the distorted PA tracks on the longitude-PA diagrams) can be
understood as the result of the V pole passage (more generally as the off-equatorial
motion of polarization state on the Poincar{\'e} sphere). 

First, while the patch is passing near the pole, the pole becomes immersed within
the patch, hence all azimuths and all PAs appear in the samples recorded
at that pulse longitude. Therefore, vertical bands of power stretch across
the full vertical range in the longitude-PA diagrams. The PA loop of
PSR B1933$+$16 
is an example of such phenomenon (Mitra et al.~2016). \nct{mra2016, i19} Other
examples can be found in Ilie (2019), eg.~PSR J1900$-$2600, see Fig.~3.64 on p.~151
therein, the near-pole passage occurs at pulse longitude $140^\circ$. 

Moreover, at a fixed pulse longitude, all data samples in direct vicinity of the V
pole are likely to have similar polarization characteristics (such as, for
example, the degree of circular polarization). Therefore, the vertical band of
spread flux (in the longitude-PA diagram) should contain a component of
radiative power which has the same polarization at any PA. 
This is expected because the local
point, ie.~the V pole on the Poincar{\'e} sphere
is multiazimuthal and contributes everywhere within the whole vertical band.

Depending on its initial position in the equatorial plane, the
moving patch (that is being rotated around some equatorial axis) follows an arc that can
pass on either side of the V pole. Let us assume that the center of some patch is passing on
the 
`western' (say `American')
side.  For an extended patch, however, some part
of the flux (in a peripheric region of the same patch) will simultaneously pass on the other
(`eastern' or `Australian') side. Passing on the other side implies that azimuth (hence PA)
will change in the opposite way (the azimuth increases on one side, while it 
decreases on the other side of pole). This is why a minor part of flux
follows the upward distortion of PA in the PA loop of B1933$+$16, whereas most of
the observed flux follows the downward path (Mitra et al.~2016).
\nct{mra2016}
Fig.~\ref{blobth} shows this effect for a Gaussian patch that 
missed the V pole by $5^\circ$.
Panel {\bf c} shows the longitude-PA diagram which results from the
continuous patch
motion shown in panel {\bf b}.\footnote{In agreement with observations (PSR
B1451$-$68, J1900$-$2600, B1933$+$16), 
the modelled OPM-transition feature is weak because of the vertical stretch of flux
on the longitude-PA diagrams. In Figs.~\ref{blobth}c and \ref{blobth2}c,
however, the
flux is normalized separately for each longitude, to make the feature 
clearly visible.}
The same V-pole-graze  
is responsible for the PA bifurcations in the model result
of Fig.~13 in Dyks (2017).

\subsection{Bifurcations of PA track}

In PSR B1237$+$25 the PA track bifurcates into two branches of comparable
magnitude. Smith et al.~(2013) \nct{srm13} have shown that the PA follows different branch
(turning either upward or downward) in different pulsation modes. The upper 
branch is followed in the normal mode, whereas the lower one -- in the
core-bright abnormal mode. As shown in Figs.~\ref{blobth} and \ref{blobth2}, 
this is exactly
what happens if the patch passes on
either side of the V pole of the Poincar{\'e} sphere. 
The difference is thus caused by the fact that the azimuth (and PA) is sampled in the opposite 
direction while passing on either side of the V pole. 

The side on which the pole is passed by changes abruptly with a small and
continuous displacement of the arc that is followed on the Poincar{\'e} sphere. 
In terms of the model
based on the coherent superposition of two waves of different amplitude, a minor change of
the amplitude ratio can switch the side on which the pole is passed. This is
why the bifurcation looks as so `drastic' phenomenon on the longitude-PA
diagram. Actually, the ratio of the 
superposed waves' amplitudes 
(consequently, the initial position
of patch on the Poincar{\'e} sphere) may change very little to produce
the opposite turn of the PA track (cf.~Figs.~\ref{blobth} and \ref{blobth2}). 

The approach of a polarization state to the V axis of the Stokes parameter space
tends to simultaneously increase $|V|/I$ and decrease $L/I$, hence the minima
of the latter coincide with maxima of $|V|/I$ in the observations. 

This interpretation of core polarization supercedes that of Dyks (2017),
where opposite changes of phase delay were considered in
different pulsation modes (section
4.7.1 therein). That model implied opposite sign of $|V|$ in both pulsation
modes. 

Here the PA track bifurcation is caused by the different side of the V pole
passage, albeit confined to the same hemisphere of the Poincar{\'e} sphere.
Therefore, this interpretation is consistent with the same sign of $V$ observed in both pulsation
modes, ie.~both in the
upper and lower branch of the PA bifurcation. In terms of the coherent wave
superposition, this implies a slightly different wave amplitude ratio in both
pulsation modes. In the case of B1237$+$25, however, 
the question of whether the side of pole
passage is changed continuously or in a discrete way remains unsolved. 

The pole-passage-driven change of PA may be caused by intrinsic emission
mechanism or propagation effects. In addition to these, the value of PA can simultaneously change because
of the RVM effect (the star-spin-induced change of the sky-projected magnetic field
direction). These two effects can easily combine to produce the range of PA variation
exceeding $180^\circ$, even in the case when the PA curve is equatorward
(or flat in the profile wings, see Mitra \& Rankin 2008). \nct{mr2008}

The pole passage effect transports the observed flux between the orthogonal PA
tracks, thus making the impression of two modes, even if there is only one.  
This implies that separation of modes into OPMs needs to be done by
following the modal patches on the Poincar{\'e} sphere, rather than following the RVM
PA tracks, which has been normal practice so far (eg.~Smith et al.~2013). 
To isolate the modes, each modal patch must be watched separately, 
regardless of which PA track it belongs to.

\section{Conclusions}

Pulsar data contain two different types of 
OPM jumps - the traditional one associated with exchange of the brighter 
mode, and another one - resulting from the V pole passage. The latter is `not
real' since it does
not involve exchange of the dominant mode nor even requires the presence  of the
second mode. A single polarization mode (single patch on the Poincar{\'e}
sphere) can be detected as two OPMs, as soon as the mode passes through the
V pole somewhere in the profile. Since the pole's
coordinates transform to a plane in the well known specific way, the
longitude-PA representation of polarization suffers from `spurious' effects which
disappear or look much less enigmatic on the Poincar{\'e} sphere. 

Whatever is the physical origin of the near-meridional circularization,
several observed complexities have its origin in just the coordinate transformation:
the misleading impression of OPM jumps, vertical spread of flux across all PAs, bifurcations of
PA track, `horns and tongue' shape of the PA distortion in B1933$+$16, the `too-large'
range of
the 
equatorward PA, and the apparent replacement of power between the observed PA tracks.

The Poincar{\'e} sphere representation of pulsar polarization is therefore
indispensable for clear data interpretation and it is much desired to
supplement the traditional viewing methods. Separation of polarization modes
should be best done on the Poincar{\'e} sphere whereas  
the traditional modelling of observed PA must allow
for transitions of a single RVM curve between the primary (strong) and secondary 
(weak) regions of PA on the longitude-PA diagram.  

 This paper has focused on effects of near-pole passage of polarization
state, typical of near-meridional circularization. 
This must be just a special case of more general
off-equatorial motion through the Poincar{\'e} sphere. Polarization data such
as those of PSR B1913$+$16 (Fig.~1 in Dyks 2017) \nct{d2017}
show maxima of $V/I$
coincident with OPMs, albeit $L/I\sim50\%$ is very high there. 
Such cases with $L/I > |V|/I$ suggest modal patch motion at a further distance from
the V pole, ie.~a motion that is oblique with respect to the equatorial
plane of the Poincar{\'e} sphere (similar to following the ecliptic instead 
of the equator in the Solar System). This corresponds to the coherent superposition of
orthogonal natural waves that are polarized elliptically, although other
interpretations are allowed. 
The phenomenon is worth further study from both theoretical and
observational side.

Aside from its unsettled physical origin, the off-equatorial 
motion of polarization state, including the
near-meridional circulation of patch on the Poincar{\'e} sphere, is the key effect that
complicates the look of observed PA curves, especially within core components.

   

\section*{acknowledgements}
This work was supported by the grant 2017/25/B/ST9/00385 of the National Science
Centre, Poland.

\bibliographystyle{mn2e}
\bibliography{listofrefs2}

\end{document}